# In-Storage Domain-Specific Acceleration for Serverless Computing

Rohan Mahapatra   Soroush Ghodrati   Byung Hoon Ahn   Sean Kinzer   Shu-Ting Wang
Hanyang Xu   Lavanya Karthikeyan   Hardik Sharma[§]   Amir Yazdanbakhsh[b]
Mohammad Alian[‡]   Hadi Esmaeilzadeh

**A**lternative **C**omputing **T**echnologies (ACT) Lab
University of California, San Diego

[§]Google     [b]Google DeepMind     [‡]University of Kansas

{rohan, soghodra, bhahn, skinzer, shutingwang, hanyang, lkarthik}@ucsd.edu   {hardiksharma, ayazdan}@google.com

alian@ku.edu     hadi@ucsd.edu

**Abstract**

While (I) serverless computing is emerging as a popular form of cloud execution, datacenters are going through major changes: (II) storage disaggregation in the system infrastructure level and (III) integration of domain-specific accelerators in the hardware level. Each of these three trends individually provide significant benefits; however, when combined the benefits diminish. On the convergence of these trends, the paper makes the observation that for serverless functions, the overhead of accessing disaggregated storage overshadows the gains from accelerators. Therefore, to benefit from all these trends in conjunction, we propose In-Storage Domain-Specific Acceleration for Serverless Computing (dubbed *DSCS-Serverless*[1]). The idea contributes a serverless model that utilizes a programmable accelerator embedded within computational storage to unlock the potential of acceleration in disaggregated datacenters. Our results with eight applications show that integrating a comparatively small accelerator within the storage (DSCS-Serverless) that fits within the storage's power constraints (25 Watts), significantly outperforms a traditional disaggregated system that utilizes NVIDIA RTX 2080 Ti GPU (250 Watts). Further, the work highlights that disaggregation, serverless model, and the limited power budget for computation in storage device require a different design than the conventional practices of integrating microprocessors and FPGAs. This insight is in contrast with current practices of designing computational storage devices that are yet to address the challenges associated with the shifts in datacenters. In comparison with two such conventional designs that use ARM cores or a Xilinx FPGA, *DSCS-Serverless* provides 3.7× and 1.7× end-to-end application speedup, 4.3× and 1.9× energy reduction, and 3.2× and 2.3× better cost efficiency, respectively.

[1]*DSCS-Serverless* is short for **D**omain-**S**pecific **C**omputational **S**torage for **S**erverless Computing.

**ACM Reference Format:**
Rohan Mahapatra, Soroush Ghodrati, Byung Hoon Ahn, Sean Kinzer, Shu-Ting Wang, Hanyang Xu, Lavanya Karthikeyan, Hardik Sharma, Amir Yazdanbakhsh, Mohammad Alian, and Hadi Esmaeilzadeh. 2024. In-Storage Domain-Specific Acceleration for Serverless Computing. In *29th ACM International Conference on Architectural Support for Programming Languages and Operating Systems, Volume 2 (ASPLOS '24), April 27-May 1, 2024, La Jolla, CA, USA.* ACM, New York, NY, USA, 19 pages. https://doi.org/10.1145/3620665.3640413

## 1 Introduction

(I) Serverless computing is emerging as a prevalent form of cloud execution that has been adopted across different market sectors such as smart transportation [1, 2], entertainment/broadcasting [3–5], e-commerce [6], fintech [7], etc. This adoption is backed by the public cloud services such as AWS Lambda [8], Google Cloud Functions [9], and Azure Serverless Computing [10]. The popularity of serverless is driven by ease of programming, pay-as-you-go pricing model, and alleviating the need for managing the cloud execution by the developers.

Besides this shift in the cloud-native application development, datacenters are going through major changes: (II) storage disaggregation in the system infrastructure level [11–17], and (III) integration of domain-specific accelerators [18–50] at the hardware architecture level. Disaggregation is enabled by the increase in network bandwidth to hundreds of Gbps and reduction in latency to single-digit microseconds [14]. Disaggregation has shown promising results in resource utilization, elasticity, and failure mitigation in datacenters [14, 51, 52]. While the improvements in networking is making storage disaggregation a viable solution, the failure of Dennard scaling [53] and the dark silicon phenomenon [54–56] has ignited a golden age of domain-specific accelerators [57]. These accelerators have made their way into the datacenters of major cloud providers including Amazon [20], Google [18, 58], Meta [21], and Microsoft [59, 60].

The trend towards serverless has coincided with these two structural changes in the infrastructure and the hardware. Each of these trends individually provide significant benefits but collectively poses challenges. On the one hand, the gains from domain-specific accelerators can potentially expand serverless usecases [61–64] and/or potentially improve their speed and efficiency. On the other hand, serverless functions operate on ephemeral data and they need to read and store their inputs and outputs from persistent storage for every invocation [15, 65–67]. To that end, we make the observation that with disaggregated storage, the overhead of moving input and output data from remote storage limits the benefits from acceleration. The gains will be limited since current accelerators are inherently designed to myopically focus on the compute and are not meant to deal with the significant data movement cost in serverless functions. Observing these insights, as shown in Figure 1, the paper explores the confluence of the three trends and devises a pathway towards maximizing the benefits from accelerators for serverless computing on disaggregated datacenters. We propose In-Storage Domain-Specific Acceleration for Serverless Computing (dubbed *DSCS-Serverless*).

This idea contributes a serverless model that leverages a relatively small programmable accelerator within storage to unlock the potential of acceleration in disaggregated datacenters. The proposed model does not advocate moving back heavy compute to the storage but takes *a more balanced approach* by integrating a rather small accelerator within the storage device to mitigate the communication overheads when applicable. These programmable accelerators are activated and utilized when a serverless function belongs to their corresponding domain. However, placing an accelerator within the storage comes with challenges.

**Tight power constraints.** According to commercial designs, storage devices adhere to stringent power budgets [68–71]. Furthermore, this power budget is apportioned between the flash and the accelerator. As such, one of the primary challenges is to architect an in-storage accelerator that not only covers a broad range of applications in the domain, but also adheres to the tight design constraints. We explore using various in-storage compute platforms (Domain-Specific Accelerator, ARM CPU, Low-power GPU, and FPGA) for a domain of serverless application while abiding by the constraints imposed by the storage. Considering the constraints, we also perform a Pareto design space exploration that examines more than 650 accelerator configurations.

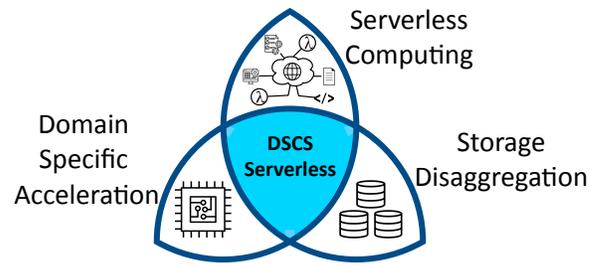

**Figure 1.** This paper devises *DSCS-Serverless* at the conjunction of three different trends in datacenters: (1) serverless computing in the programming level; (2) storage disaggregation in the system infrastructure level; and (3) domain-specific accelerators in the hardware level.

**In-storage accelerator integration with serverless system stack.** Serverless functions use frameworks such as OpenFaaS [72] and Kubernetes [73] for deployment and orchestration. The challenge is how to minimally change the serverless system stack to integrate *DSCS-Serverless* without disruption to traditional application operations within a disaggregated datacenter. Our solution enables serverless functions to be offloaded to the in-storage accelerator seamlessly using software hints provided at function deployment time. We also design an OpenCL device driver that enables serverless functions to access the accelerator. Furthermore, we handle the scheduling of both data and functions and the challenges associated with storage scaling, data replication, and fault tolerance in a disaggregated datacenter that leverages **D**omain-**S**pecific **C**omputational **S**torage **Drive** (*DSCS-Drive*).

We choose machine learning/neural networks as the domain to design a programmable accelerator and showcase an implementation of *DSCS-Serverless*. We evaluate the system through a rigorous study with eight real-world, latency-critical, *end-to-end* serverless applications inspired from AWS Lambda case studies [74–81]. Evaluations show that integrating a comparatively small accelerator for *DSCS-Serverless* significantly outperforms a traditional disaggregated system that utilizes the NVIDIA RTX 2080 Ti GPU. In comparison, *DSCS-Serverless* achieves 2.7× end-to-end speedup, 4.2× energy reduction, and 3.0× better cost efficiency. *DSCS-Serverless* also performs better than existing computational storage solutions that either use microprocessors [69, 82] (3.7× end-to-end speedup, 4.3× energy reduction, and 3.2× better cost efficiency) or FPGAs [70] (1.7× end-to-end application speedup, 1.9× energy reduction, and 2.3× better cost efficiency).

To put it in a nutshell, the paper contributes:

- ***The insight that overhead of moving data from remote storage*** limits the benefits from acceleration for serverless functions in disaggregated datacenters.
- ***The* DSCS-Serverless *execution model*** that leverages a relatively small programmable accelerator within the storage device to accelerate a domain of serverless functions.
- ***A software stack that seamlessly integrates in-storage accelerator with existing serverless models,*** handling storage-specific challenges such as data placement, scalability, and utilization, along with serverless considerations such as function placement, scalability, and cold starts.

- ***The insight that disaggregation, serverless model, and the limited power budget for computation in storage require an alternative design*** *than the conventional practices of integrating microprocessors and FPGAs within storage.*

## 2 Background and Motivation

Serverless computing is a cloud computing model that allows developers to write and run code without worrying about the underlying infrastructure, scaling, or billing.

**Why Serverless? A motivating use-case.** Wildfires pose a serious threat to California's flora and fauna, environment, and infrastructure. San Diego Gas & Electric (SDG&E), an energy services company in Southern California, uses drones to capture images of forest and uploads them to the cloud, as shown in Figure 2. An object detection serverless application hosted on AWS analyzes the images for potential fire hazards in real-time [81, 83], enabling SDG&E to respond effectively and swiftly to wildfire risk. We use this application below to describe a model serverless execution flow.

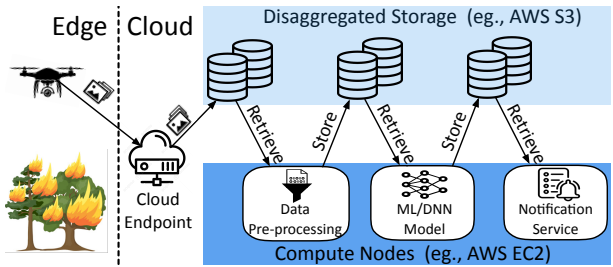

**Figure 2.** A serverless computing workflow for an object detection application that detects wildfires using drones. The drones capture images of the forest and send them to the cloud, where a serverless application consisting of three functions analyzes the images to detect potential fire. The functions uses disaggregated storage for data exchange.

### 2.1 Life of Serverless Application

**Deployment.** Figure 2 depicts a widely used serverless pipeline for object detection [76] that consists of three functions: Data Pre-processing, Machine Learning/Neural Network (ML/DNN) Inference, and Notification Service. During deployment, the applications are modeled as serverless functions and chained together using RESTful APIs. The application developer (in this case SDG&E) also configures metadata constraints (timeout, trigger mechanisms, hardware requirements, etc.) into a configuration file (eg., YAML in AWS) for each of the functions. Since serverless functions are stateless, SDG&E also allocates a persistent storage (such as AWS S3) that is used by the functions to retrieve and store data. SDG&E then deploys the application to a cloud serverless provider such as AWS Lambda [8], Google Cloud Functions [9], etc.

**Invocation.** The *Remote Sensing* application is invoked when data (eg., image) is sent from the drone to the cloud datacenter where the *Remote Sensing* application is deployed. The data arrives at the storage that was configured during deployment by SDG&E. Based on the function's deployment constraints outlined in the YAML file, the serverless framework running on the cloud (AWS Lambda, OpenFaaS, etc.) launches the function on a compute node. The function then retrieves the data using an RPC from the storage node as shown in Figure 2. At the storage node, this RPC invokes a series of system calls to read data from the physical storage over PCIe. The data are then serialized [58], converted to network packets and transmitted to the compute node. After function execution, the output data (ephemeral or not) are stored back to the persistent storage following similar steps discussed above for reading the data. Moreover, if the function utilizes a specialized domain-specific accelerator (DSA) such as GPUs, ASICs or FPGA at the compute node, the compute node has to further initiate a data transfer (e.g. cudaMemcpy for GPUs) to the DSA devices' memory generally over PCIe [50, 61]. Overall, these steps are expensive for serverless functions that come with strict Service Level Objective requirements [67, 84] since they involve RPCs [65, 85], system calls [86], and I/Os [67].

### 2.2 Characterization of Serverless Applications

As demonstrated above, there are variegated components that contribute to the end-to-end latency of a serverless application. We profile serverless applications (Table 1) on AWS EC2 instances using the methodology described in Section 6.1 to understand how these components contribute to the end-to-end application latency.

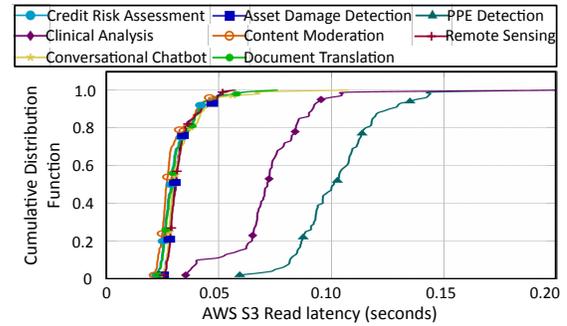

**Figure 3.** Cumulative distribution function for reading inputs from AWS S3 for different benchmarks.

**Communication in disaggregated storage datacenters.** Figure 3 shows the cumulative distribution function for reading data from remote S3 storage across a range of benchmarks (refer Table 1). *The results show that accessing storage suffer from tail latency.* The average latency difference between the median and the $99^{th}$ percentile is a factor of 110% for read accesses. This long tail latency is primarily because of remote storage that increases the network communication overhead. Our analyses about tail latency of serverless functions is commensurate with prior studies [65, 87, 88]. Indeed, recent work has devised solutions to mitigate network latency for microservices or serverless functions through RPC accelerations [58, 85, 89], specialized network protocol for RPC [90], and communication bypassing/fused functions [66, 67].

**Computation vs. communication.** Figure 4 shows the compute, communication (network + I/O), and the system stack overhead of launching the function using OpenFaaS and Kubernetes. We observe that latency to access the remote storage accounts for a significant portion of the end-to-end application runtime (on average > 55%).

*The average latency for reading and writing the data to the remote storage is greater than the time it takes to perform the computation.* In fact, Credit Risk Assessment, Asset Damage Detection, and Content Moderation consists of ≥ 70% communication. This communication overhead is naturally expected because of the serverless function execution flows discussed earlier (§2.1).

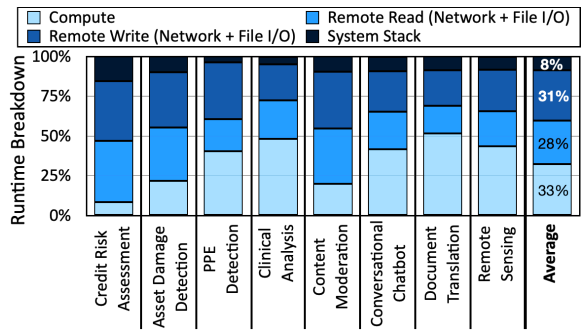

**Figure 4.** Runtime latency breakdown for application modeled as serverless functions deployed on AWS EC2 with remote S3 storage.

**Domain-specific acceleration.** Accelerators have been integrated into major cloud providers including Google [18, 58], Amazon [20], Meta [21], and Microsoft [60]. The efficiency of these accelerators can unlock additional usecases in serverless computing [61, 63, 64]. However, the primary target of these accelerators is commonly focused on computation efficiency. Figure 4 shows that the maximum speedup attainable by accelerating the compute is capped at 1.52×. This is because *with remote storage, the overhead of moving input and output data (on average > 55%) limits the benefits from acceleration in serverless applications.* As such, the overall benefits of the current paradigm of acceleration for serverless computing in disaggregated storage datacenters is strictly limited by the Amdahl's Law [91].

## 3 Overview of DSCS-Serverless

*DSCS-Serverless* is an execution model for serverless computing that integrates small programmable accelerators within some of the storage drives in disaggregated datacenters. The model aims to leverage in-storage accelerator to reduce the communication overheads, without shifting high-performance compute devices (eg., GPUs) back to the storage node. The accelerators are designed for a specific domain of functions and provide acceleration when applicable.

Figure 5 (a) shows how *DSCS-Serverless* fits into the current datacenter rack infrastructure. Specifically, in the *DSCS-Serverless* model, we replace some of the existing storage drives with computation storage drives that house an in-storage domain-specific accelerator (DSA). We term this type of storage device as **D**omain-**S**pecific **C**omputational **S**torage **Drive** (dubbed *DSCS-Drive*).

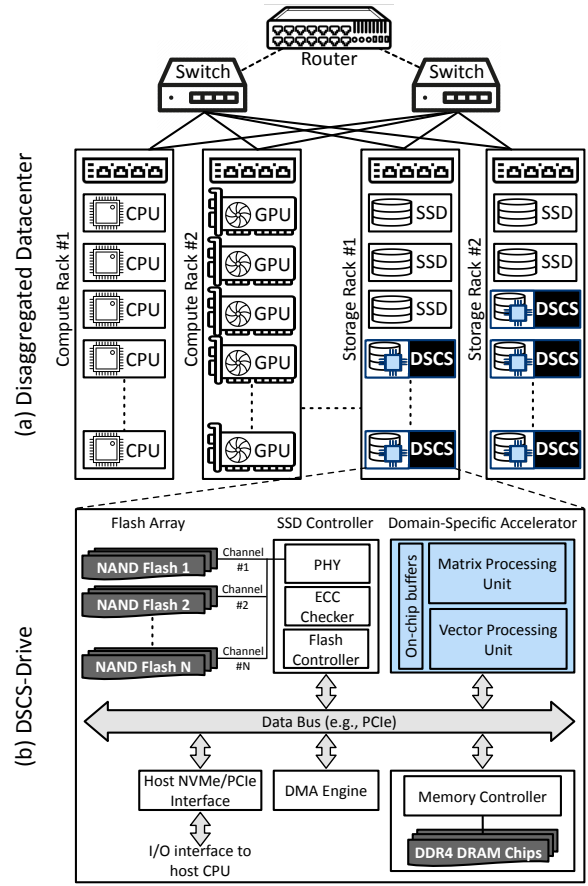

**Figure 5.** System architecture of *DSCS-Serverless*. (a) How *DSCS-Serverless* fits into the datacenter rack infrastructure. (b) The architecture of *DSCS-Drive* that integrates a DSA next to the flash array.

**DSCS-Drive.** Figure 5 (b) shows the architecture of *DSCS-Drive*. *DSCS-Drive* integrates a relatively small in-storage DSA on which a domain of functions execute. The DSA within the *DSCS-Drive* directly communicates with the flash array (e.g. NAND flash, SLC, MLC, SSD, etc.) using the dedicated data bus (eg., PCIe peer-2-peer links [70]). The *DSCS-Drive* also houses a small DRAM memory that acts as a staging buffer to exchange data between the host CPU, the flash array, and the DSA. The DMA engine is used to transfer data between each of these components.

With such system architecture, serverless functions that are amenable to acceleration using the in-storage DSA obviate the extravagant network and I/O data transfer overheads (remote read/write parts in Figure 4). Furthermore, *DSCS-Serverless* does not consume CPU cycles in the storage node, except for initiating the data transfer to the DSA located near the storage node. Thus, it does not interfere with the applications that run concurrently on the storage node's CPU. *DSCS-Serverless* also improves the resource utilization of the storage node by enhancing the storage drives with an additional computational capability without hampering the conventional storage functionality to serve applications.

### 3.1 Life of a Serverless Function in *DSCS-Serverless*

We contrast the life of a serverless function in *DSCS-Serverless* and the traditional serverless execution model.

① ↦ In contrast to traditional system, in *DSCS-Serverless*, a function that is amenable to acceleration is directly deployed to *DSCS-Drive* that has the data. This thereby eliminates the costly invocation of a compute node. In §5.3, we describe the details of how *DSCS-Serverless* identifies the functions amenable to acceleration.

② ↦ Instead of fetching data from remote storage via costly RPC requests, *DSCS-Serverless* employs its driver to initiate a peer-to-peer (P2P) data transfer from flash array to DRAM memory within the *DSCS-Drive*. As described in §2.2, each function invocation in the traditional system confers a notable data read/write latency. Specifically, in the traditional system an AWS S3 read request is translated into a RPC that incurs the network latency to access the remote storage. Upon reaching the storage node, the request further requires a protobuf deserialization and a *read* system call to access the data over PCIe interface. In fact, to reduce the high cost of protobuf operations, prior work has proposed hardware accelerators [58]. In contrast, *DSCS-Serverless* circumvents the costly protobuf operations by performing a single system calls that initiates a P2P data transfer from the flash array to the DRAM memory bypassing the host's software stack. Once the data are entirely transferred to the DRAM memory, the execution of the serverless function starts.

③ ↦ After the function execution completes, the DSA sends an interrupt over PCIe to the host CPU to initiate a P2P transfer of the results to the flash array. Once the transfer completes, the host may invoke subsequent serverless function calls. This is in contrast to the traditional system in which data transfer over network and I/O events occur to write the results to persistent storage.

## 4 Architecture Design for DSCS-Serverless

In this section, we first discuss the architecture of the programmable domain-specific accelerator (DSA). Then, we showcase a methodology to derive the optimal DSA configuration under tight storage power constraints. We also demonstrate a technology scaling analysis to project the performance for more recent technology nodes. While we present the design space exploration framework for a single domain of applications, it can be readily employed to find optimal DSAs for arbitrary application domains.

**Architecture for machine learning/neural networks applications.** As per an IBM survey, machine learning/neural network (ML/DNN) including transformers such as large language models inference applications are one of the fastest growing domains in serverless computing, accounting for more than 40% of services [10, 92, 93]. As such, we set the primary target of our design to accelerate such applications. The important design decisions that we considered for the architecture are: programmability such that it can cater to a wide range of commonly deployed ML/DNN applications and low-power such that it can abide by the stringent power constraint imposed by the storage [68, 70, 94]. Specifically, the architecture should support a wide range tasks such as image classification, object detection, semantic segmentation, linear/logistic regression, neural machine translation, conversational AI, generative AI, data pre-processing, etc.

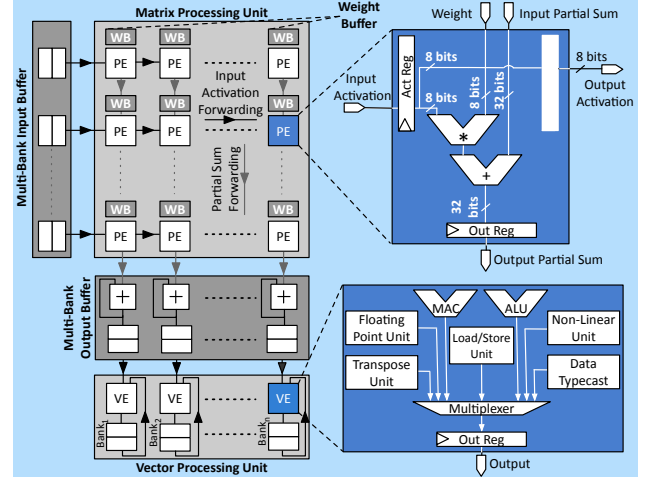

**Figure 6.** Architecture of the in-storage domain-specific accelerator (DSA) for machine learning/neural networks within the *DSCS-Drive*.

We analyze the applications listed in Table 1 to identify the various types of operations that are present. We find that apart from the General Matrix Multiplication (GeMM) operations (matrix multiplication, convolution, etc.), the ML/DNN models used in the applications also consists of operations such as element-wise mathematical operations (add, sub, multiply, etc.), element-wise activation functions (ReLU, GeLU, TanH, Sigmoid, etc.), data layout transformations (eg., reshape, transpose, etc.), reduction-based operations (layer-normalization, batch normalization, mean, etc.), and data type conversion operations (eg., fp32 to fp16). To this end, as illustrated in Figure 6(b), we design a DSA that consists of a *Matrix Processing Unit* (*MPU*) to execute the GeMM layers and tightly couple it with a *Vector Processing Unit* (*VPU*) to execute the other layers described above.

### 4.1 Domain-Specific Accelerator Microarchitecture

**Matrix Processing Unit.** Figure 6 shows the *MPU* and the microarchitecture of a processing engine (PE) inside the *MPU*.

The *MPU* consists of a 2D array of Processing Elements (PEs) and dedicated multi-bank buffers (scratchpad) for input activations, weights, and outputs as shown in Figure 6. Each bank of the buffer unit is exclusively shared across PEs within a row. The execution flow of such architecture is similar to conventional systolic-array accelerators [24, 31, 95]. At each cycle input activation tensors are fetched from input buffers and shared across the PE units within a row. The partial sum from the PEs are forwarded in a waterfall fashion per column. Once the computations across the array of PEs conclude, the results are either fed to the *VPU* for ensuing operations or written back to DRAM.

**Vector Processing Unit.** Figure 6 shows the microarchiecture of the *VPU* and vector engine (VE) that performs the computations. The *VPU* is a Single Instruction Multiple Data (SIMD) architecture designed primarily to execute activation functions (e.g. Relu, LeakyRelu, Tanh, Sigmoid), pooling, quantization, vector arithmetic computations, and datatype casting which are prevalent in emerging data analytic workloads [49, 96]. The *Vector Processing Unit* and the *Matrix Processing Unit* are closely connected through the shared multi-bank output buffer, as Figure 6 shows. This way, the *VPU* can access the data from the *MPU* directly, without having to get it from the DRAM. Furthermore, ML/DNN applications also require data pre-processing/post-processing such as tokenization, normalization, scaling, and datatype casting [75, 78, 97, 98], etc. These transformations are commonly packaged as separate serverless functions [80, 97]. We utilize the *VPU* to execute these data pre/post-processing functions as well thereby expanding the type of functions that can leverage the DSA.

### 4.2 Design-Space Exploration for Optimal In-Storage Domain-Specific Accelerator

Design-space exploration is crucial to ensure that DSA architecture adheres by the tight power and thermal constraints of storage drives. In this section, we discuss the power and thermal constraints imposed by the storage drives and then describe the details for the design-space exploration.

**Power and thermal constraints of storage drives.** Storage drives have tight design constraints primarily because of the limited PCIe power budget ($\leq 25$ watts [68]) that is the exclusive source of power for these units. To provide an estimate of the storage drive's limited power budget, commercially available computational storage drives such as the Samsung's SmartSSD [70, 71] have merely a TDP (ideal) of 25 watts. Furthermore, the power source in these computational storage drives is shared between the flash array and compute (eg., FPGA) [69–71]. As such, performing design-space exploration for DSA is crucial.

**Design-space exploration objective.** There are various conflicting factors to take into account to identify optimal DSA design points for storage drives. Commensurate with prior work [55, 99], we use throughput (frames per second/tokens per second) as the performance metric. Capital expense incurred by ASIC fabrication is another determining factor for designing data centers [100, 101]. However, measuring the precise capital expense is not pragmatic and kept confidential by major cloud providers. Therefore, we use chip area as a proxy for the ASIC fabrication cost. We use DSA's power consumption to assess the feasibility of a design point, which is capped by the storage drive's limited power budget. The objective of design-space exploration is to find design points that are on the Pareto frontier of the power-performance and area-performance. Particularly, we only select DSA architecture configurations that are on the Pareto frontiers while abiding by the tight power and area constraints of storage drive. We choose the open source 45 nm FreePDK [102] technology node for our baseline analysis in the design-space exploration.

**FPGA implementation and simulation methodology.** To effectively support the design space exploration of DSA, we designate the number of PEs, systolic array X-Y dimension, on-chip buffer sizes, and memory bandwidth as configurable parameters. We then implement and synthesize DSA using the methodology detailed in §6.1. Since hardware simulations for the entire design points ($>650$) is not practical, we develop a cycle-accurate simulator to closely model the latency and power of our designed DSA (refer §6.1). For power and area numbers, we use the synthesized values using 45 nm technology node. We followed the methodology in [103] to scale the results to 14 nm, which is relatively similar to the technology node of Samsung SmardSSD [71]. We provide the details of methodology in Section 6.1.

**Domain-specific accelerator search space.** We use Google TPUv1 [18] with 256×256 PEs, 28 MB of on-chip buffers, and 34 GB/s memory bandwidth as the standard design point. We then scale this design by varying the number of PEs from 4×4 to 1024×1024 with a power of 2 stride. We proportionally scale the buffers to provide sufficient on-chip resources for PEs. However, we set the maximum total buffer size to 32 MB because large buffer sizes significantly increase the power consumption, exceeding the tight power constraints for DSA. We use three realistic memory bandwidth in the search space, namely DDR4 (19.2 GB/s), DDR5 (38 GB/s) and HBM2 (460 GB/s).

**Pareto-optimal design points.** Figure 7 demonstrate the power-performance and Figure 8 shows the area-performance results across a range of design points. The curved lines (Pareto frontier) shows the best power-performance and area-performance tradeoffs for DSA configurations. By definition, Pareto frontier provides the most optimal points in a design space. We exclude the design points that are either infeasible because of design constraints or significantly inefficient in terms of throughput. The power and throughput of each design point is an average across set of target benchmarks (See Table 1). The design point at the bottom left portion of Figure 7 (power-performance), represents a 4x4 systolic array with 128 KB on-chip buffer and DDR4 memory. The top right design configuration shows a 128×128 systolic array with 4 MB on-chip buffer and DDR5 memory (denoted by Dim128-4MB in Figure 7).

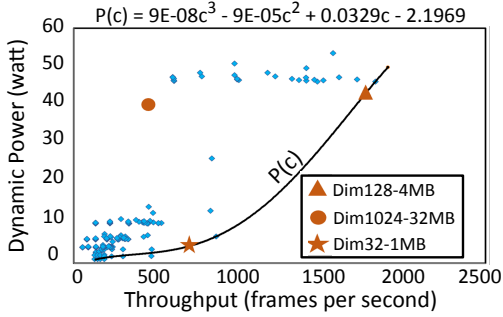

Figure 7. Power-Performance frontiers, 45 nm Tech Node.

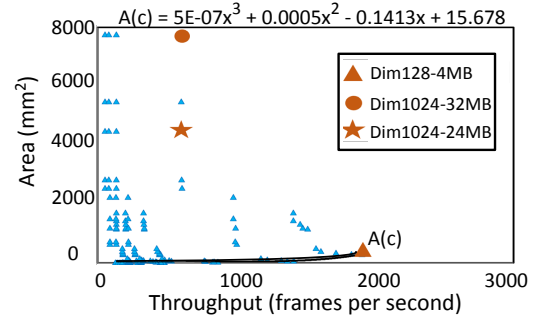

Figure 8. Area-Performance frontiers, 45 nm Tech Node.

The results of our design-space exploration indicates that a 1024×1024 systolic array delivers significantly lower throughput compared with 128×128 array. This is because the DSA employs a tiling-based execution mechanism. For a batch size of one, a 1024×1024 systolic array does *not* show the most optimal performance because the compiler (Section 5.1) aims to obtain the optimal tiling such that the DSA overlaps the memory transfers for a tile with the computation of preceding tile. If the tile sizes are large, the cycles spent on memory transfer outweigh the compute cycles. This is similar to CPU pipelines where stalls reduce the IPC. Through our design space exploration, we find the most optimal DSA configuration to be a 128×128 systolic array with 4 MB on-chip scratchpad, and DDR5 memory.

## 5 In-Storage Accelerator Integration with Serverless System Stack

*DSCS-Serverless* integrates in-storage accelerators with serverless computing framework by addressing system software, storage, and serverless specific considerations. In achieving this goal, *DSCS-Serverless* strives to introduce minimal changes to the existing software/framework avoiding disruptions to traditional application operations in a disaggregated storage datacenter.

### 5.1 Integration of DSCS-Serverless with System Software

This section discusses the challenges and solutions of adapting *DSCS-Serverless* into the existing system software, such as modifying the programming model, the compiler, the device drivers, etc. The goal is to make *DSCS-Serverless* easy to use without adding any extra work for the application developer.

**System software stack.** *DSCS-Serverless* is deployed atop OpenFaaS, an open source serverless framework. OpenFaaS is deployed on Kubernetes to orchestrate containers, and uses Promethus [104] for cluster monitoring and telemetry. *DSCS-Serverless* can also be readily deployed into other serverless platforms such as Apache OpenWhisk [105].

**Programming model.** In serverless, developers define their applications as a Directed Acyclic Graph (DAG) of decoupled functions [67]. During deployment of these functions, the developer provides a configuration file (eg., YAML) that describe the properties and constraints of the function (eg. dependencies, timeout, access mechanism, storage, etc.). *DSCS-Serverless* extends this YAML file to enable developers to mark in-storage DSA acceleratable functions. In addition, developers provide a container that packages the accelerated serverless function with the appropriate device drivers and libraries.

**Device driver and libraries.** To support in-storage acceleration, *DSCS-Serverless* includes an OpenCL device driver. The driver implements standard interfaces for mapping storage space to physical address space of both the storage node and the DSA's configuration registers and memory. Additionally, the driver orchestrates direct P2P data transfer between the storage and the DSA that bypass the storage node's system stack and utilizes dedicated PCIe links. The OpenCL driver also abides by the OS security checks for access control to both storage and DSA. The acceleratable function's container include all dependent libraries such as OpenCL framework and associated runtime/tools that is needed by the DSA.

**Compiler support.** We develop a compilation stack capable of code generation for different DSA configurations and for a range of evaluated machine learning/neural networks benchmarks. The functions that utilized the DSA are implemented using Pytorch and stored as ONNX (Open Neural Network Exchange) files. The front-end part of the compiler performs a range of optimizations [106], including operator fusion to minimize off-chip data movement. Then the compiler performs DSA design configuration (e.g. number of PEs, memory bandwidth) specific optimizations such as padding and tiling to maximize the DSA's utilization. Once these optimization passes complete, the compiler generates the hardware configuration specific optimized executable code. This code is packaged along with the serverless function in the container. Note that we rely on the developer to partition their application into server functions that can or cannot be accelerated using a in-storage DSA.

### 5.2 Storage Node Considerations for DSCS-Serverless

In this section, we present how to integrate *DSCS-Serverless* into the storage servers. We address three key challenges: how to place the data intelligently on the storage drives near the accelerator, how to transfer data seamlessly between the accelerator, storage drive, and the host CPU, and how to ensure scalability of DSCS in disaggregated datacenters.

**Data placement.** To enable effective utilization of in-storage accelerators, it is essential that the data to be processed is situated on the same storage drive that houses the accelerator. Cloud service providers today offer various storage classes [107] for different types of data (hot and frequently accessed data, cold data, and archived data, etc.). *DSCS-Serverless* resembles the baseline system that uses a disaggregated key-value storage setup, similar to AWS S3. This setup involves replicating data across multiple storage nodes to ensure data reliability. If a serverless function is accelerable, *DSCS-Serverless* maps one of its replicas to a *DSCS-Drive* (a new class of storage). Storage nodes typically divide data into fixed-size chunks (ranging from 1MB to 64MB [108]) before storing them on physical media. However, *DSCS-Serverless* assumes requests (data) are not distributed across multiple devices within the same storage node due to the small size ($\leq$ 20MB in AWS S3 [109]) of serverless requests. In those exceptional instances where data is indeed distributed across multiple storage drives, *DSCS-Serverless* has the flexibility to either revert to default CPU execution or execute data in parallel across multiple CSDs.

During invocation, all data requests for a function designated as accelerable during deployment are directed to these *DSCS-Serverless* capable storage drives. As the number of requests (data) increases, it is possible to store different requests on separate drives that support *DSCS-Serverless*. This is because requests (data) are independent of each other, and therefore, the scheduling of requests can be distributed across different drives. The scheduler relies on Prometheus telemetry to decide whether to employ in-storage acceleration or execute the function in a conventional manner depending on if the node is busy.

**Storage utilization.** The accelerator in *DSCS-Serverless* is an optional extra capability to utilize the storage more. It does not affect conventional serverless or storage functionality, as it can be bypassed for normal storage operations.

**Storage scalability.** One of the design goals of disaggregation is the independent scalability of resources. For example, compute resources can be scaled independently from remote storage (such as S3), which offers virtually unlimited storage capacity. *DSCS-Serverless* does not compromise this since *DSCS-Serverless* capable nodes can also function as conventional storage node for applications that do not need compute acceleration capabilities. *DSCS-Serverless* scales horizontally by adding more compute capability drives. Hence, *DSCS-Serverless* does not limit independent storage scalability.

**Host and storage communication.** The DSA and the flash storage device in the computational storage use the same PCIe links to communicate with the host node. A switch in the computational storage routes the requests to either the flash storage device or the DSA, based on the request type. There is also a dedicated P2P connection between the flash storage device and the DSA for fast data transfer. *DSCS-Serverless* uses this P2P connection to communicate with the flash storage device over PCIe links, bypassing the host CPU.

### 5.3 Serverless Function Considerations for DSCS-Serverless

*DSCS-Serverless* is a system that uses a specialized hardware accelerator to run serverless functions on the storage nodes that contain the data. This requires the functions to be assigned to the same node that has the data and the accelerator. We also need to examine how *DSCS-Serverless* enables serverless-specific features such as fail-over support, function chaining, and cold start. In this section, we describe how we support these serverless-specific properties.

**Function scheduling.** We extend the centralized Kubernetes scheduler to expose storage nodes that can utilize in-storage accelerators and map accelerable serverless functions execution to such nodes if the data resides on the node. The scheduler uses a simple First Come First Serve (FCFS) scheduling policy for the incoming requests, leveraging Prometheus to monitor availability and prevent overloading a single Kubernetes pod (in this case the storage node). Note that similar to traditional serverless [8–10], a function instance on the accelerators also does *not* support preemption and follows a run-to-completion execution policy. Once a function is offloaded for computation on *DSCS-Serverless*, the storage node marks its compute status as *busy*. The scheduler does not offload more functions until the node becomes available.

**Future directions for optimized scheduling for DSCS-Serverless.** Previous works have explored various scheduling optimizations for different applications, both serverless and non-serverless [66, 67, 110–115]. However, none of them have considered the *DSCS-Serverless* execution model, which enables in-storage acceleration of serverless functions. Applying these optimizations to *DSCS-Serverless* is a potential future direction that can improve performance. For instance, scheduling functions based on their criticality and importance can enhance the performance of *DSCS-Serverless* by assigning long-running functions to nodes that support *DSCS-Serverless*. Likewise, scheduling policies that consider the whole serverless application DAG and use *DSCS-Serverless* for applications that have many acceleratable functions can also boost the performance. However, scheduling techniques that depend on task heterogeneity and affinity to different accelerators may not be effective, as *DSCS-Serverless* already knows which functions can be accelerated at deployment time.

**Fail-over support.** In the event of DSA unavailability within the storage, *DSCS-Serverless* may be unable to process a function. The scheduler then defaults to conventional execution using remote compute nodes (e.g., CPU) for function execution (§2.1). This is possible as *DSCS-Drive* can operate as standard storage drives supporting storage APIs (e.g., AWS S3 APIs [107]). We utilize existing Kubernetes mechanisms for fail-over and container migration, leveraging telemetry (via Prometheus) for node health monitoring.

**Function chaining.** *DSCS-Serverless* maps chained functions to the same *DSCS-Drive* that has the data if they can be accelerated by the same DSA. If not, the function falls back to CPU. Additionally, *DSCS-Serverless* handles stateless functions that don't write to shared data structures, simplifying function scheduling and duplication.

**Cold starts.** Functions in *DSCS-Serverless* incur the same cold start as functions in traditional platforms. A function experiences cold start when the function's container image is pulled from a remote registry, unpacked, and has to pass a health check. This happens when a function is deployed for the first time to a node or when function replicas are created by increasing the number of nodes (horizontal scaling) from $N$ to $N+1$. Similar to the conventional serverless execution mechanism where the function is kept warm on the compute node's memory for a certain period of time, *DSCS-Serverless* also stores the function on the DSA's memory for some duration preemptively waiting for new requests. In case when another different function is scheduled on the DSA, instead of evicting the old function, the DSA offloads the function's container image to the flash storage using the P2P interconnect. Next time the same function is scheduled on the same storage node, *DSCS-Serverless* can just use the P2P to load the function instead of fetching it over the network from the serverless framework such as OpenFaaS which store the container image in their registry.

## 6 Evaluation
### 6.1 Methodology
**Benchmarks.** To evaluate the efficacy of *DSCS-Serverless*, we use eight real-world latency critical machine learning or neural network including large language models (LLMs) applications representing serverless pipelines deployed on AWS Lambda [74–81]. Table 1 shows the suite of applications, their description, serverless functions, the machine learning/neural network model, the number of parameters, and the corresponding inputs/outputs sizes. Since the exact models used in AWS Lambda functions are not publicly available for some benchmarks, we use representative and state-of-the-art inference models from Hugging Face [116] that provide similar functionality (e.g. we use ResNet-50 [117] for AWS Rekognition [76] that offers image classification). We containerize all the serverless functions by using OpenFaaS.

**Baseline system setup.** For the Baseline (CPU), we use *Amazon EC2 c5.4xlarge* instance with *Intel® Xeon® Platinum 8275CL* CPU and use an *IAM* account to connect the instance to a S3 object storage in the same region. The EC2 instances run *Ubuntu 20.04.4 LTS* with kernel version *5.13.0-1029-aws*. We launch a Kubernetes cluster on the EC2 instance and deploy OpenFaaS on a pod [73]. During deployment phase, applications are enlisted in the OpenFaaS function registry.

**Evaluation of compute platforms.** Table 2 lists the specifications of all the evaluated platforms. The *Traditional Platforms* that are currently utilized for serverless function deployment, where the compute (consisting of Baseline (CPU), GPU, or FPGA) accesses the remote storage via the network. We also evaluate *Conventional In-Storage/Near-Storage platforms (NS)*, where the compute is within the storage. Since these platforms are not available in datacenters, we set up the infrastructure locally similar to the baseline setup described above. We consider three low power near-storage platforms: quad-core ARM CPU [118] (denoted by NS-ARM), a low-power Nvidia Jetson TX2 mobile GPU [118] (denoted by NS-Mobile-GPU), and Samsung SmartSSD that houses an FPGA [71] (denoted by NS-FPGA). Since we did not have access to ARM Cortex A53 used in commercial CSDs [69, 82], we use a more powerful ARM core (Cortex A-57) for our evaluation. Upon invocation, for Baseline (CPU), each function within the application is launched on a Kubernetes pod running on the CPU. On all platforms, we use the available compute unit such as the ARM CPUs, GPU, FPGA, or DSA to execute both the data pre-processing (Function 1) and ML/DNN model inference (Function 2) for each application. Function 3 always runs on a CPU in a compute node.

**System performance measurements.** For the Baseline (CPU) measurement, we use the aforementioned baseline setup on AWS EC2 instance and invoke the application by generating 10,000 sequential requests using hey [119], an open-sourced http load generator to measure the latency. We use the 95[th] percentile latency for all our analyses similar to prior work [66, 67]. To measure the latency for all other *traditional compute platforms*, we create containers with the required environments (e.g. ONNX Runtime for GPU or Xilinx XRT for FPGA in addition to their corresponding drivers) that access the remote storage via the host CPU.

**Table 1.** Benchmarks, their brief description, interconnected chain of serverless functions, machine learning/neural network model, parameters, and input/output dimension.

| Application | Description | Interconnected Serverless Functions | | | DNN Model | Number of Parameters | Input/Output Dimensions |
|---|---|---|---|---|---|---|---|
| | | Function 1 | Function 2 | Function 3 | | | |
| Credit Risk Assessment | Identify positive and negative credit risks for loan approval | Normalization | Logistic Regression | Notification | Logistic Regression | 200 | (200)/ (1) |
| Asset Damage Detection | Detect damages to objects using images captured by CCTV | Image Preprocessing | Image Classification | Notification | ResNet-50 | 25 million | (3,224,224)/ (1,1000) |
| PPE Detection | Detect workers' safety gear in factories to prevent hazards | Image Preprocessing | Object Detection | Notification | YOLOv3 | 65 million | (3,416,416)/ (255,52,52) |
| Clinical Analysis | Identify salient elements in medical scans | Image Preprocessing | Semantic Segmentation | Notification | FCN | 54 million | (3,224,224)/ (3,224,224) |
| Content Moderation | Detect inappropriate or offensive content | Image Preprocessing | Image Classification | Notification | EfficientNet | 11.5 million | (3,227,227)/ (1000) |
| Conversational Chatbot | Answers input text questions | Tokenization | Question & Answering | Response | BERT-Base | 110 million | (128,768)/ (128,768) |
| Document Translation | Localize website contents for cross-lingual communication | Tokenization | Neural Machine Translation | Response | GPT-2 | 1.5 billion | (1,128)/ (1,128) |
| Remote Sensing | Infrastructure monitoring using drones | Image Preprocessing | Scene Classification | Notification | Vision Transformer | 632 million | (3,224,224)/ (1000) |

**Table 2.** Specification of the traditional, conventional near-storage, and proposed platforms used for evaluation.

| | TRADITIONAL PLATFORMS | | | CONVENTIONAL NEAR-STORAGE PLATFORMS (NS) | | | PROPOSED |
|---|---|---|---|---|---|---|---|
| | Baseline (CPU) | GPU | FPGA | NS-ARM | NS-Mobile-GPU | NS-FPGA | DSCS-Serverless |
| Chip | Xeon 8275CL | RTX 2080 Ti | Xilinx Alveo U280 | Arm Cortex A57 | Pascal GPU | Samsung SmartSSD | Domain-Specific Accelerator |
| Cores/ PEs | 16 cores | 4352 CUDA Cores | 1024 PEs | 4 cores | 256 CUDA cores | 256 PEs | 16,384 PEs |
| Memory | 32 GB | 11 GB GDDR6 | 32 GB(2 MB on-chip) | 4 GB | 4 GB | 4 GB (2 MB on-chip) | 4 GB (4 MB on-chip) |
| TDP | 240 W | 250 W | 225 W | 15 W | 15 W | 18 W | 4.2 W |
| Frequency | 3 GHz | 1.35 GHz | 250 MHz | 2 GHz | 1.3 GHz | 250 MHz | 1 GHz |
| Technology Node | 14 nm | 12 nm | 16 nm | N/A | N/A | 16 nm | 14 nm |
| PCIe | N/A | Gen3x16 | Gen4x8 | N/A | N/A | Gen3x4 | Gen3x4 |

For *DSCS-Serverless*, we utilize Samsung SmartSSD to implement the DSA configuration specified in Table 2 and utilize the OpenCL driver (§5.1) to measure the end-to-end execution time, that encompasses the P2P read/write data transfer latencies and computation latency. Additionally, we include the system software overhead by incorporating it from the Baseline (CPU) into the end-to-end latency of *DSCS-Serverless*. We execute each application 10,000 times on the Samsung SmartSSD and sample the 95$^{th}$ percentile latency. For the case of *Conventional Near-Storage platforms*, we develop an analytical model where we replace the DSA compute latency measured for the *DSCS-Serverless* system with the respective compute latency of NS-ARM, NS-Mobile-GPU, or NS-FPGA.

**Hardware implementation and synthesis.** We implement the DSA in 15k lines of Verilog and synthesize it using Synopsys Design Compiler R-2020.09-SP4 with FreePDK 45nm standard cell library. The design achieved a 1GHz frequency. To synthesize the DSA for Samsung Xilinx SmartSSD FPGA, we use the Xilinx Vitis/Vivado toolchain. We also use the Xilinx Vivado to obtain the resource utilization, timing, power, and thermal statistics for the FPGA analysis.

**Simulation infrastructure.** We compile each machine learning/neural network model to the domain-specific accelerator's ISA and generate executable binaries. We develop a cycle-accurate simulator for DSA's ASIC implementation, which uses compiler-generated instructions, and provides cycle counts and energy statistics. We compare the simulator results with the FPGA implementation of DSA on the Samsung SmartSSD for the same design configuration and frequency to verify the closeness of the cycles by an error margin of $\leq 10\%$. We use this simulator to obtain the performance/energy numbers for DSA ASIC implementation and design projections mentioned in §3.

To evaluate *DSCS-Serverless* at scale under high request arrival rates, we develop a simulation infrastructure that models a datacenter rack. We assume the maximum number of compute platforms (*DSCS-Serverless* or Baseline (CPU) with remote storage) available on a data center system is 200. The simulator also has a scheduler with a queue (depth 10,000) that holds incoming requests that cannot be executed on a node. The scheduler handles the incoming requests using the policy described in §5.3. Similar to prior work [120], we also generate an application trace (Figure 13 (a)) by randomly sampling functions from the benchmarks (Table 1) using Poisson distribution and impose load on the system for 20 minutes. We use this setup to measure the wall clock time (§6.2.2), which is the cumulative wall clock time the platforms take to process all the incoming requests from the application trace. We also use this setup to measure the effect of cold starts (Figure 17).

**Power measurement.** We measure the compute, PCIe and system stack power dissipation and combine them to report the energy efficiency of the system. Although, serverless systems also use the network (Ethernet/Internet), measuring the power for it was not feasible and therefore we omit the network power for all the traditional systems. We use the Intel RAPL[121] and MSR registers to get the Baseline (CPU) power. To obtain the power for the NS-ARM and NS-Mobile-GPU, we use the *NVPModel* tool from NVIDIA Jetson TX2 Development Kit [118]. We use Xilinx Vivado to measure the power for FPGA implementation of DSA on Samsung SmartSSD and Alveo u280. To obtain power for the ASIC DSA, we use synthesis results to measure the logic cell power and CACTI-P [122] to model on-chip memory energy. For PCIe, we use the per-bit PCIe power reported in prior work [123].

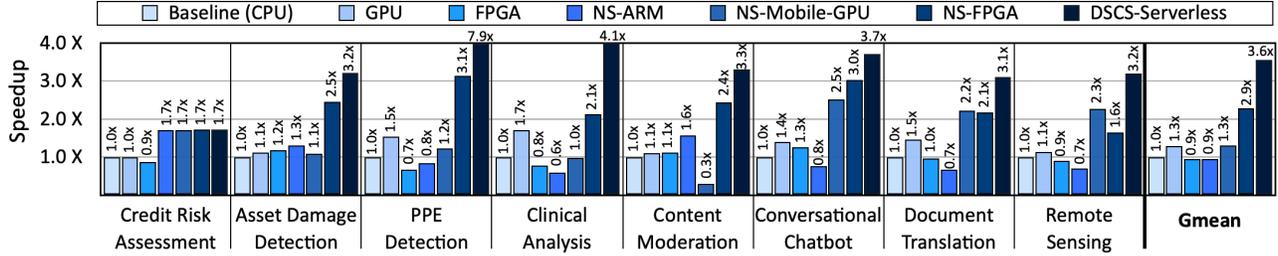

**Figure 9.** Normalized speedup for applications designed as serverless functions.

**Cost efficiency model.** To assess if a new design offers cost savings over other systems, we evaluate cost efficiency using the methodology in prior work [101], which is the average peak throughput over total cost and time of ownership as shown in the equation below.

$$\text{Cost Efficiency} = \frac{\text{Throughput} \times T}{\text{CAPEX} + \text{OPEX}}$$

The total cost is composed of two key components: CAPEX and OPEX. *CAPEX* represents the initial capital expenditure, encompassing the entire system's cost, including processing units, networking, compute servers, and storage. We use market prices for off-the-shelf components [18, 69, 71, 118, 124–127] and estimate ASIC's $ cost using the analytical model from ASIC Clouds [100]. OPEX covers the ongoing operating costs (over a three year period with 30% utilization rate) and consists of power expenses for processing units, network transfer, storage, and cooling similar to E3 [101]. It is the product of the power (watts) for various components in the cluster, the time for which the cluster is active (T) and the average industrial electricity rate in the U.S. ($0.0975/kWh) in 2023 [128].

### 6.2 Experimental Results

#### 6.2.1 Single Node Evaluations.
In this section, we evaluate the performance, runtime breakdown, and energy reduction of *DSCS-Serverless* on a single node with various benchmarks.

**Performance comparison with traditional serverless platforms.** Figure 9 compares the performance of various traditional platforms (with remote storage) listed in Table 2 across all studied benchmarks. The speedups are normalized to the Baseline (CPU) that is commonly used for serverless execution in public clouds. On average, *DSCS-Serverless* provides 3.6× speedup over the baseline across all benchmarks. *DSCS-Serverless* also outperforms GPU with remote storage by 2.7×. This is because, first, the inherent data movement latency to remote storage limits the performance benefits from the high-end GPU. Second, using batch size one for serverless scenarios causes underutilization in GPUs. Utilizing FPGA with remote storage exhibits a slight performance dip compared to the Baseline (CPU). This is attributed to the constrained resources of the FPGA for implementing a high performance DSA, coupled with the driver overhead associated with the FPGA.

> For the applications we studied, results show a lightweight in-storage accelerator (4.2 watts) outperforms a high-end GPU (250 watts) with remote storage. This is because the data movement overhead from remote storage limits the acceleration benefits in disaggregated datacenters.

**Performance comparison with conventional in-storage platforms.** To tackle the communication overheads, we also analyze various in-storage computing platforms. Figure 9 compares the performance (normalized to Baseline (CPU)) of various conventional in-storage (denoted as NS) scenarios across all studied benchmarks. As shown in Figure 9, NS-ARM which utilizes a general-purpose compute platform (quad-core ARM CPUs) within the storage slightly underperforms compared to the Baseline (CPU). Using specialized accelerators such as NS-Mobile-GPU provides 1.35× speedup while leveraging NS-FPGA unlocks 2.2× speedup. The speedup for low-power NS-FPGA seems counter-intuitive compared to high-power FPGA (with remote storage) because the latter was bottlenecked by the communication overhead. *This analysis shows that the overhead of moving input and output data from remote storage limits the benefits from acceleration.* Nevertheless, NS-FPGA's performance is still bounded by its limited resources and low frequency.

As shown in Figure 9, *leveraging a domain-specific architecture within the storage (*DSCS-Serverless*) unlocks additional*

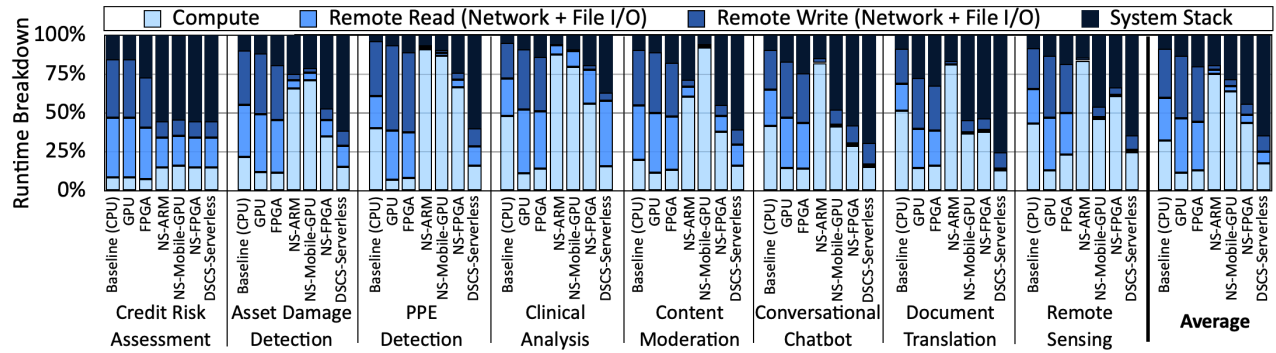

**Figure 10.** Normalized runtime breakdown.

benefits and provides 3.7× and 1.7× speedups over the conventional approaches of using microprocessors (NS-ARM) and FPGAs (NS-FPGA) in the storage, respectively. *Credit Risk Assessment* shows the least speedup because logistic regression is not computationally intensive while *PPE Detection* achieves the maximum speedup because moving compute to in-storage reduces the significant data movement that the benchmark otherwise incurs. In general, *DSCS-Serverless* offers performance benefits for functions characterized by an intensity in both computational and communication aspects, while functions with low compute intensity can still benefit from a general-purpose in-storage processor (e.g., *Credit Risk Assessment* that uses logistic-regression in Figure 10).

> To address the challenges at the confluence of infrastructure disaggregation, serverless computing, and storage power constraints, the results advocate for domain-specific accelerators within storage, departing from conventional approaches of integrating CPUs and FPGA.

**Runtime breakdown analysis.** Figure 10 shows the runtime breakdown across the individual system components for the benchmarks and platforms. We see that for *traditional platforms* with GPU/FPGA (with remote storage), the compute portion is significantly reduced due to hardware acceleration. However, the data transfer over the network limits the effective speedup achieved by the hardware acceleration. This significant data transfer is addressed by the in-storage platforms where moving the compute closer to storage reduces the data movement, shifting the bottleneck back to the compute. The DSA further accelerates this compute portion unlocking additional performance gains. Overall, we observe that leveraging *DSCS-Serverless* shifts the bottleneck from the compute and communication to other components such as the system stack.

For benchmark *Credit Risk Assessment*, Figure 4 shows that data movement accounts for approximately 75% of the runtime. Intuitively, moving compute to in-storage should provide at least a 3× speedup. However, we observe a 1.8× speedup because of two reasons. First, as mentioned in the methodology 6.1, *function 3* is launched on the CPU and experiences the network and IO latency similar to traditional systems. Second, the latency incurred due to the in-storage driver reduces the theoretical speedup. As depicted in Figure 10, for *DSCS-Serverless* the bottleneck now is the latency incurred by the *function 3* to read the data from persistent storage and the system stack overheads.

**Energy reduction comparison.** Figure 11 analyzes the end-to-end system energy reduction achieved by *DSCS-Serverless*. On average, *DSCS-Serverless* provides 3.5× energy reduction over the Baseline (CPU) system and 1.9× reduction over the NS-FPGA (Samsung SmartSSD), the most competitive baseline. FPGAs have significantly higher static energy dissipation and thus cannot match the energy efficiency on an ASIC. Although leveraging DSA provides significant energy reduction (29× over Baseline (CPU)), the total system energy reduction is bounded by the system stack and *f3* function being executed on the CPU. The trends in energy reduction are similar to the speedup, with *PPE Detection* showing the maximum gains (8×) and *Credit Risk Assessment* showing the minimum (1×).

#### 6.2.2 At Scale Evaluations.
This section assesses the cost efficiency of *DSCS-Serverless* when integrated into a large-scale datacenter setting. It also shows how *DSCS-Serverless* can handle concurrent applications and lower latency than Baseline (CPU) on a large scale with a substantial number of requests.

**Cost efficiency.** Figure 12 shows the cost efficiency for various platforms normalized to the Baseline (CPU). Results show *DSCS-Serverless* offers the highest cost efficiency (3.4×) compared to the Baseline (CPU), while NS-FPGA (Samsung SmartSSD) ranks second (1.6×). This result is intuitive, since over the initial period of usage, the *CAPEX* cost of building hardware is dominant. As time goes on, the *OPEX* cost, that is cost of operating (electricity cost) becomes more dominant. Since *DSCS-Serverless* consumes less energy compared to other platforms, its cost efficiency increases over time.

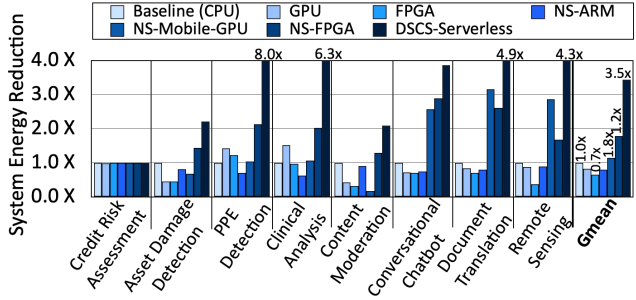

**Figure 11.** Normalized system energy reduction for application.

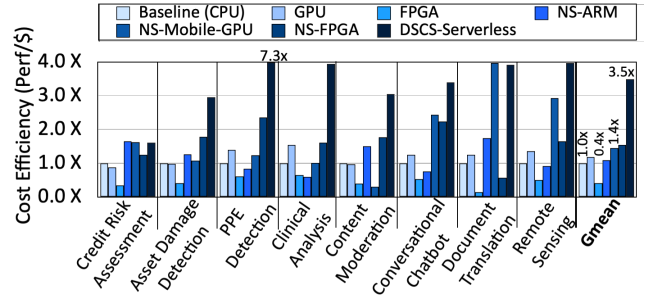

**Figure 12.** Normalized cost efficiencies.

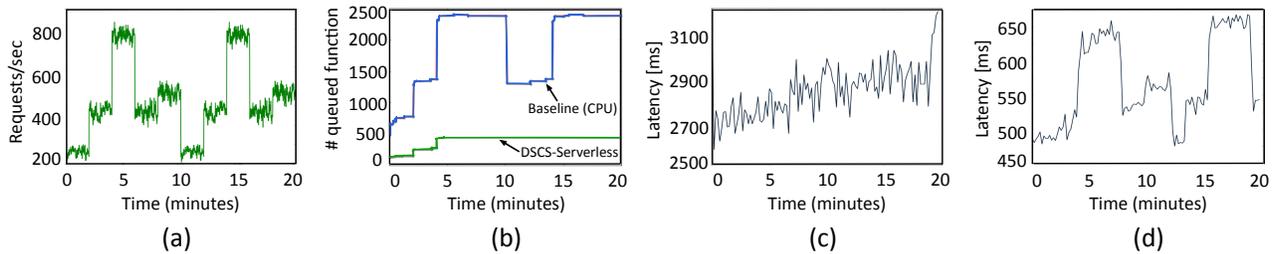

**Figure 13.** (a) Input trace. (b) Number of concurrent functions at the queue. Latency (c) Baseline CPU (d) *DSCS-Serverless* latency.

**Number of concurrent function invocations.** Figure 13 (a) shows the synthetic input trace of requests with different inter-arrival rates; specifically with bursty arrival patterns similar to prior work [120]. Figure 13 (b) illustrates the number of concurrent functions that are at the scheduled queue of both *DSCS-Serverless* and Baseline (CPU) under different load conditions. *DSCS-Serverless* has a lesser number of functions compared to the Baseline (CPU) due to its near-storage acceleration, which reduces latency thereby enabling a single function instance to service a larger number of requests. Moreover, both systems exhibit a delayed response to the decline in input requests because functions persist in the memory for some time even after the load decreases. However, this is more pronounced for the Baseline (CPU) because it has more function invocations that can handle requests less efficiently. On the contrary, *DSCS-Serverless* improves the throughput of the system since each *DSCS-Serverless* instance can process more requests per second as compared to the baseline.

**Wall clock latency comparison.** Serverless systems scale horizontally by replicating function instances on demand to handle additional requests and improve performance and availability. However, developers often set a maximum number of function instances for their applications to control the cost. We set the maximum number of function instances to 200 for both *DSCS-Serverless* and Baseline (CPU). We use the synthetic workload shown in Figure 13 (a) to simulate different request patterns and measure the latency to evaluate *DSCS-Serverless*. Figure 13 (c) and Figure 13 (d) show the wall clock latency of both systems using the input load shown in Figure 13 (a). The Baseline (CPU) shows a steady increase in latency with time. This is because the baseline has higher request processing latency since it not only has to move data from remote storage to memory but also cannot accelerate the workload. This means that the baseline system accumulates more and more requests in the scheduler's queue, which increases the latency of request processing. One way to improve the baseline would be to increase the number of function instances, which in turn would incur additional costs. *DSCS-Serverless* on the other hand achieves scalability and low wall clock latency by processing larger number of requests efficiently at each node level (*DSCS-Drive*).

> For the applications trace we studied, *DSCS-Serverless* improves cost efficiency (3.4×) and reduces wall clock latency latency in large-scale evaluations, demonstrating its economic and performance advantages.

**6.2.3 Sensitivity Analysis.** In this section, we conduct a sensitivity study to evaluate the impact of different factors on the performance of *DSCS-Serverless* compared to Baseline (CPU), such as batch size, tail latency effect, increasing number of accelerated functions, and cold start.

**Batch size.** Figure 14 shows the sensitivity of the *DSCS-Serverless* end-to-end performance with respect to batch size (refer 1). We sweep the batch size from one to 64 across all benchmarks and report the latency of *DSCS-Serverless* normalized to the Baseline (CPU) with remote storage, using the same batch size. The rationale behind limiting the batch size to 64 is that AWS Lambda has a strict cap on the network payload size for serverless functions [109]. Relative to the Baseline (CPU), the performance improvements of *DSCS-Serverless* increase from 3.6× for batch size 1 to 15.8× for batch size 64. This performance improvement stems from (1) reducing the communication overheads of transferring batched data to the compute node and (2) the capability of the DSA in reusing the weights across the batch, thereby improving the computation significantly. The improvements are more pronounced for *Conversational Chatbot* and *Document Translation*, since these benchmarks deploy language models with a large number of weights, where *DSCS-Serverless* leverages batching to amortize the cost of loading weights by reusing them across the input batch.

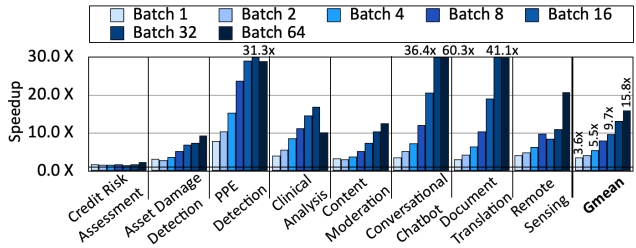

**Figure 14.** Sensitivity to batch size.

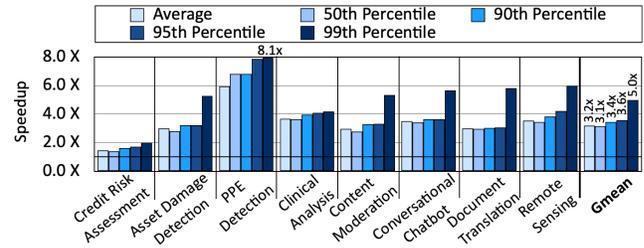

**Figure 15.** Sensitivity to storage access latency.

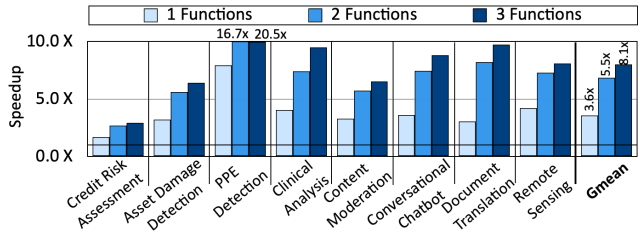

**Figure 16.** Sensitivity to the number of accelerated functions.

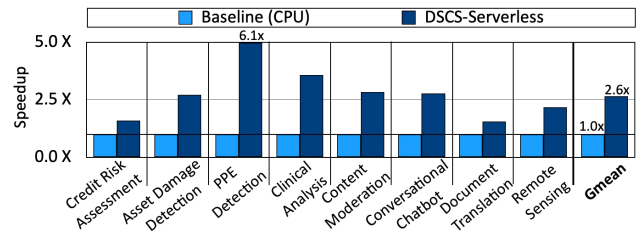

**Figure 17.** Sensitivity to cold vs. warm containers.

**Tail latency effect.** Accessing remote storage can incur long tail latency (§2). To understand this variability and its implications on *DSCS-Serverless* performance, we perform a sweep across various latency distributions for the PCIe, P2P, and network. Figure 15 shows the implications of tail latency normalized to Baseline (CPU) with the same latency distribution. Results suggest *DSCS-Serverless* is robust to network and I/O tail latency since it removes the data movement over them. On average, *DSCS-Serverless* provides 5.0× speedup for the $99^{th}$ percentile and 3.1× speedup for $50^{th}$ percentile.

**Number of accelerated functions.** To analyze the sensitivity of the *DSCS-Serverless* to the number of accelerated functions, we create synthetic benchmarks by adding either one, two, or three additional accelerated functions to the application. These functions are duplicates from the original benchmarks' *function 2*. The label in Figure 16 refers to the number of replicated functions and performance is normalized to the Baseline (CPU) running the same function configuration. Results show that by increasing the number of functions that are offloaded to *DSCS-Serverless*, the improvements escalate (from

3.6× to 8.1×). This is because it emulates the scenarios in which the serverless applications are composed of more complex pipelines with multiple functions [129, 130]. Using these complex pipelines would incur more pronounced computation and communication overheads to the end-to-end execution, both are addressed significantly by domain-specialization and near-storage computation of *DSCS-Serverless*.

**Cold start.** Figure 17 shows the speedup of *DSCS-Serverless* over Baseline (CPU). Both *DSCS-Serverless* and the baseline use cold containers where they pull the container image (including the weights for the model) and load it to the memory of the DSA. Since the models are large, the time to load a model accounts for a significant portion of the end-to-end latency, thereby reducing the speedup from 3.6× to 2.6×. However, as mentioned in Section 5.3, cold latency is incurred by both *DSCS-Serverless* and the baseline systems. Further, only the first invocation incurs a cold latency while all subsequent invocations can potentially hide the cold latency using preemptive horizontal scaling (Refer Section 5.3).

## 7 Related Work

Individually, the emergence of serverless computing, the shift towards storage disaggregation, and the adoption of domain-specific accelerators has provided significant benefits but collectively they pose interesting challenges. The paper explores the confluence of the three trends and provides a pathway to unlock the true benefits from accelerators for serverless computing in disaggregated datacenters.

**Serverless and storage.** Serverless functions are stateless and ephemeral [65–67, 84]. They use persistent storage to transfer intermediate data between functions. Locus [131] focused on deriving an optimal combination of storage and fast in-memory caching while SONIC [132] used local and remote storage to pass data between functions. Pocket [15] proposed a storage system to allocate different storage resources depending on workloads to reduce cost. NumPyWren [133] identified appropriate block size to remote storage for serverless linear algebra. Jiffy [134] used in-memory caching on remote servers to accommodate intermediate data. However, it still incurs the network latency to remote storage. These papers are orthogonal to our work since they consider multi-tier storage and the possibility of efficient data passing between functions. *DSCS-Serverless* introduces a novel model of serverless computing by leveraging near-storage DSA to reduce the data movement and unlock additional benefits from acceleration.

**Systems for serverless functions.** Various systems have been proposed that optimize the performance for serverless functions. SmartNICs have been used to accelerate serverless functions [135]. Speedo [136] placed the function dispatcher on SmartNIC to avoid latency overhead. Dagger [85] accelerated RPCs using FPGA-based NIC. BlastFunction [137] exposes FPGAs to serverless framework for acceleration while Shredder [138] executes programs on CPU in the storage controller. Molecule [61] and Hardless [64] propose runtimes to enable hardware accelerators for serverless. HiveMind [139] proposes a hardware-software solution for serverless edge swarms. Overall, these solutions either enable data-movement-aware acceleration or compute-focused acceleration using GPUs/FPGAs. *DSCS-Serverless* on the other hand, leverages the insight that serverless functions are stateless and require remote storage in a disaggregated datacenter to devise a comprehensive, cross-stack near-storage serverless acceleration solution.

**Near-storage acceleration.** Prior works have explored near-data ASICs for various domains demanding large amount of data transfer. [47, 140–151]. There are commercially available products such as Eideticom's NoLoad [152] for transparent compression, Samsung SmartSSD for utilities (encryption, compression, etc.) [70], and NGD system's Newport for encryption on ARM cores [69, 82]. Deepstore [142] introduces a microarchitecture tailored for in-storage processing of DNNs and delves into SSD parallelism methods. In contrast, *DSCS-Serverless* performs an extensive design-space analysis of using various in-storage compute platforms while abiding by the constraints imposed by the storage and identifies an optimal DSA configuration to unlock the potential for acceleration of serverless functions in disaggregated datacenters.

## 8 Conclusion

Emergence of serverless computing coupled with disaggregation and hardware specialization introduces unique challenges and opportunities that emanate the overhead of communicating data from remote storage. To address this issue, the paper devises a serverless execution model that integrates a domain-specific accelerator within the storage device. Evaluation with a diverse set of benchmarks against variety of compute platforms shows significant gains in terms of performance, energy, and cost efficiency. As such, this paper marks an initial step towards utilizing accelerators for serverless execution in disaggregated datacenters.

## Acknowledgement

We thank the anonymous reviewers for their valuable comments. We thank our shepherd, Jian Huang, for his feedback and encouragement. This work was in part supported by generous gifts from Google, Microsoft, Samsung, Qualcomm, AMD Xilinx as well as the National Science Foundation (NSF) awards CCF#2107598, CNS#1822273, National Institute of Health (NIH) award #R01EB028350, Defense Advanced Research Project Agency (DARPA) under agreement number #HR0011-18-C-0020, and Semiconductor Research Corporation (SRC) award #2021-AH-3039. The U.S. Government is authorized to reproduce and distribute reprints for Governmental purposes not withstanding any copyright notation thereon. The views and conclusions contained herein are those of the authors and should not be interpreted as representing the official policies or endorsements, either expressed or implied of Google, Qualcomm, Microsoft, Xilinx, Samsung, NSF, SRC, NIH, DARPA or the U.S. Government.